\documentclass[preprint2,twoside]{hwo}

\usepackage{lipsum}

\input{hwo.h}

\begin{document}

\title{\textbf{\LARGE The ocean worlds science case for the Pollux spectropolarimeter}}
\author {\textbf{\large V. Hue$^{a,b}$, 
B. Benmahi$^{a,b,c}$, 
M. Barthelemy$^{d}$, 
T. M. Becker$^{e}$, 
J.-C. Bouret$^{a}$, 
R. J. Cartwright$^{f}$,
J.-Y. Chaufray$^{g}$, 
L. Fossati$^{h}$,  
G. Giono$^{h}$,
O. Poch$^{i}$,
U. Raut$^{e}$, 
L. Roth$^{j}$}}
\affil{$^a$\small\it Aix-Marseille Université, CNRS, CNES, LAM, Marseille, France \email{vincent.hue@lam.fr}}
\affil{$^b$\small\it Aix-Marseille Université, Institut Origines, Marseille, France}
\affil{$^c$\small\it Laboratory of Atmospheric and Planetary Physics, STAR Institute, University of Liège Belgium}
\affil{$^d$\small\it Univ. Grenoble Alpes, CNRS, IPAG, 38000 Grenoble, France}
\affil{$^e$\small\it Southwest Research Institute, San Antonio, TX, USA}
\affil{$^f$\small\it Johns Hopkins University Applied Physics Laboratory, Laurel, Maryland, USA}
\affil{$^g$\small\it LATMOS-IPSL, UVSQ Paris Saclay, Sorbonne Universite, CNRS, France}
\affil{$^h$\small\it Space Research Institute, Austrian Academy of Sciences, Schmiedlstrasse 6, 8042, Graz, Austria}
\affil{$^i$\small\it Univ. Grenoble Alpes, CNRS, IPAG, 38000 Grenoble, France}
\affil{$^j$\small\it School of Electrical Engineering, Royal Institute of Technology KTH, Stockholm, Sweden}

\begin{abstract}
  Pollux is a candidate European instrument contribution to the Habitable Worlds Observatory (HWO), designed to advance our understanding of the formation and evolution of cosmic structures in the universe, and specifically search signs of life on extrasolar planets. This high-resolution spectrograph (R\,$>$\,40,000) with polarimetric capabilities offers nearly continuous and simultaneous coverage from the FUV ($\sim$100\,nm) to the NIR ($\sim$1.9\,$\micron$), making it a versatile tool for a wide range of scientific investigations from solar system studies to cosmology. Several Solar System ocean worlds have been the focal point of the scientific community to understand the conditions of their internal saline oceans, as well as the possible emergence of life beyond Earth. The ocean world science case will leverage Pollux's UV spectropolarimetric capabilities to investigate surface reflectance and composition, characterize airglow emissions in the environments of giant-planet moons, as well as constrain the microphysical properties of atmospheric aerosols.
\\
\\
\end{abstract}

\vspace{2cm}

\section{Introduction}

The Habitable Worlds Observatory (HWO) is currently designed as a ``great observatory'', following the legacy of the Hubble Space Telescope (HST), the Spitzer Space Telescope, and the James Webb Space Telescope (JWST), to serve the broader astrophysics community. Its primary science driver is the search for life beyond the Solar System through the detection of habitable-zone rocky planets and the characterization of their atmospheres. Pollux, a proposed European contribution to HWO, shares this dual vision by enabling a wide range of science cases, from Solar System science to exoplanet and astrophysics.

\section{The Pollux spectrograph}

\begin{figure*}[ht!]
    \centering
    \includegraphics[width=0.6\textwidth]{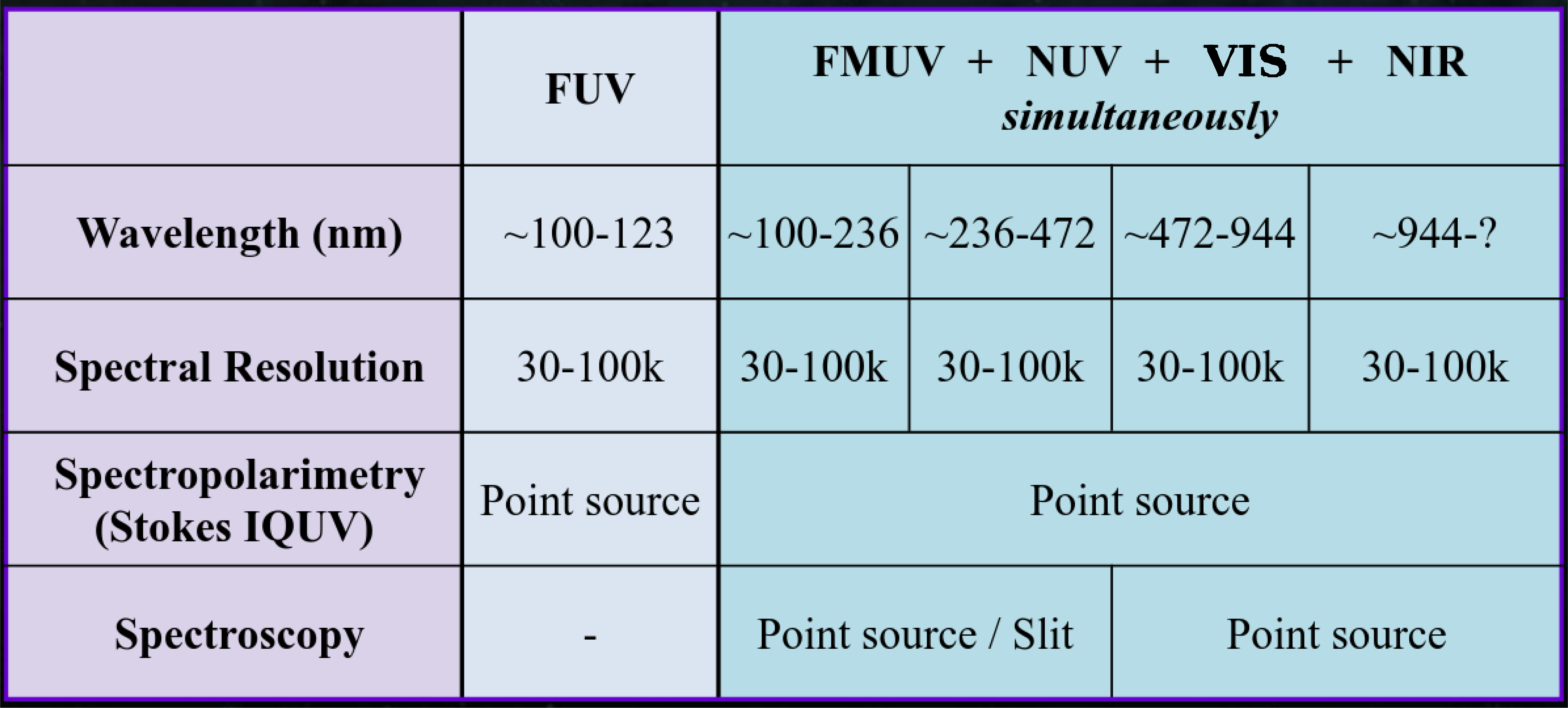}
    \caption{Summary of the current properties of the Pollux instrument. The spectral resolution is defined as $\lambda$/$\Delta \lambda$. Point-source spectroscopy traditionally involves injecting the light into a narrow aperture, while slit spectroscopy is optimized for spatially-extended sources.}
    \label{fig:summary_table}
\end{figure*}

In its current design, Pollux consists of five-high-spectral resolution classical echelle spectrographs, each covering a distinct wavelength range: Far-Ultraviolet (FUV: 100-123\,nm), Far- to Mid-Ultraviolet (FMUV: 101-236\,nm), Near-Ultraviolet (NUV: 234-472\,nm), Visible (VIS: 472-944\,nm), and Near-Infrared (NIR: 944-1888\,nm). The FMUV, NUV, VIS, and NIR arms are equipped with retractable polarimeters. All arms are initially designed for a spectral resolution of $\sim$100,000, except FMUV, which achieves $\sim$75,000. Figure~\ref{fig:summary_table} summarizes Pollux’s current specifications.

\begin{figure*}[ht!]
    \centering
    \includegraphics[width=0.9\textwidth]{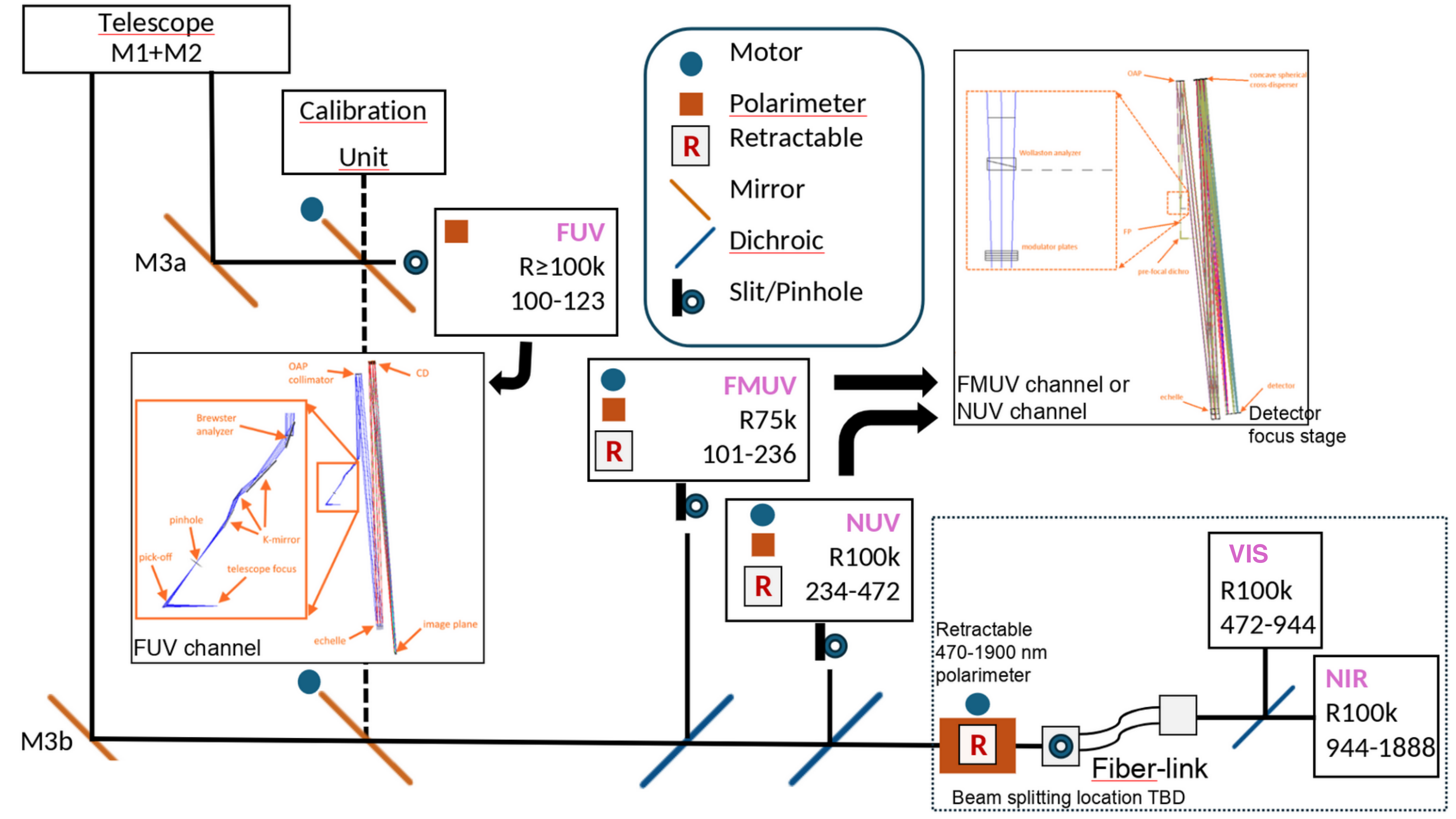}
    \caption{Pollux instrument concept schematic diagram.}
    \label{fig:Diagramm_Pollux}
\end{figure*}

The FUV arm, dedicated exclusively to spectropolarimetry, operates independently of the others. To maximize throughput by reducing reflections, the Pollux team is investigating with the HWO team the possibility of feeding it directly via a dedicated pick-off mirror (M3) after the primary (M1) and secondary (M2). The remaining arms are fed through a separate, dedicated M3 and a cascade of three dichroics that split the light to allow simultaneous operation of the FMUV, NUV, VIS, and NIR spectrographs (see Fig. \ref{fig:Diagramm_Pollux}). The manufacturability of the first dichroic, splitting at $\sim$235 nm, is currently under industrial study, at the time of writing. The FMUV and NUV arms support both pinhole and slit spectroscopy enabled by an exchangeable entrance mechanism. The FUV, FMUV, NUV, and VIS arms use large CMOS detectors, while the NIR arm relies on a JWST-like large H2RG detector. To simplify the architecture, microchannel plates are not considered at the time of writing, and all detectors operate without active cooling.
 
In September 2025, the Pollux Project Office (PO) issued a call for science cases to identify the capabilities that would best maximize the instrument scientific impact. Based on this input, the team will refine the baseline design, followed by detailed trade-off analyses. The current design therefore remains preliminary, though several features are expected to be fixed: high-resolution spectroscopy (R\,$>$\,40,000), spectropolarimetry with retractable polarimeters, and simultaneous broad wavelength coverage. 

This document describes the ocean worlds science cases enabled by the Pollux instrument, and specifically regarding its polarimetric capabilities. More general details on the assessment of the ocean worlds characterization with HWO can be found in previous publication \citep{Cartwright2025}.


\section{The Pollux ocean worlds science case}

Several ocean worlds have been identified within the Solar System that likely harbor internal saline oceans (see, e.g., \citealt{Hendrix2019}). These worlds are defined as planetary bodies that currently have known sub-surface oceans and conduits to their surfaces in some cases, or are retained at depth with limited or no communication between the ocean and surface. Confirmed ocean worlds include the jovian moons Europa and Ganymede, and possibly Callisto, as well as Saturn's moons Enceladus and Titan. Results from the Dawn mission indicate that Ceres also exhibits characteristics consistent with past ocean world activity \citep[see, e.g.,][]{castillo-rogez_future_2020}. Other candidate icy moons that may be ocean worlds include Ariel, Dione, Mimas, Pluto, Rhea, Titania, Triton.

More specifically, Europa and Enceladus display evidence of material exchange between their internal oceans and surfaces. Enceladus is known for erupting material from its subsurface ocean through active geysers, including water vapor and ice grains that contain salts and organic compounds (e.g., \citealt{Porco2006, Waite2009, Postberg2009, Postberg2018}). The surface of Europa exhibits evidence for exchange with its interior through the distribution of likely ocean-derived compounds, such as NaCl and CO$_2$, obtained with HST and JWST \citep{Schmidt2011, Becker2024, Trumbo2022, Villanueva2023, Cartwright2025_PSJ_JWST}. Furthermore, HST and Galileo magnetic and plasma wave measurements indicate transient plumes at Europa that may originate from endogenous activity \citep[e.g.,][]{Roth2014, Jia2018, Sparks2017}.

On the other hand, Ganymede and Callisto are thought to host thicker ice shells, with no recent, ocean-drived material exchange between their interiors and surfaces. Additional worlds such as Ariel, Ceres, Dione, Pluto, and Triton, are suspected to host deep reservoirs of water, based on spectroscopic, gravitational, and geological evidence \citep[e.g.,][]{Hendrix2019, castillo2023compositions}.

The existence of ocean worlds, with potentially habitable interiors, makes them important targets of study in astrobiology. The combination of abundant liquid water in their interiors, coupled with internal energy sources (e.g., tidal and/or radiogenic heating), and the presence of essential chemical building blocks, constitutes a potential subsurface habitat. NASA’s Roadmap to Ocean Worlds accordingly calls for the exploration of these bodies to \textit{characterize their oceans, evaluate their habitability, search for life, and ultimately understand any life we find} \citep{Hendrix2019}.

NASA's planned Habitable Worlds Observatory (HWO) mission would allow for unprecedented studies of ocean worlds. Some of the main cases regarding the observations and characterization of ocean worlds by HWO have been discussed in \cite{Cartwright2025}, which emphasized the utility of a UV–visible integral field spectrograph (IFS) on HWO to allow for detailed studies of potentially habitable icy moons. With adequate instrument design and specifications, HWO would allow the capture of sporadic geyser plumes from some ocean worlds, as well as to measure the spectral signatures of surface deposits enrich in ocean material (\textit{e.g.} H$_2$O, salts, organics).

Key performance requirements identified by \cite{Cartwright2025} include broad UV–visible spectral coverage (extending roughly 90–800\,nm at minimum) with high spectral resolving power (R$\sim$5,000) and high sensitivity (SNR$>$100), combined with a spatial resolution on the order of 0.04'' in order to resolve surface regions and plume sources (Figure \ref{fig:HWO_res})

\cite{Cartwright2025} also recommended a multi-epoch observational campaign that would allow long-term monitoring to track secular exospheric/surface compositional changes, as well as track transient events such as geyser activity. Such an instrument would be expected to enable discoveries, e.g., by continuously monitoring geyser eruptions and interior–surface exchanges that are key for investigating the ocean-world characteristics.

We present here additional cases that would be leveraged by the spectropolarimetric capabilities such as those provided by the Pollux instrument. HWO–Pollux would be capable of making key science observations of confirmed and candidate ocean worlds, by:

\begin{enumerate}
\item Probing the ocean worlds surface properties through spatially-resolved spectropolarimetric measurements of the surface reflection
\item Probing electromagnetic and atmospheres environments of the ocean worlds through polarimetric measurements of their local airglow emission and scattered light
\end{enumerate}

We further describe each of these points.


\subsection{Ocean worlds surface properties}

When sunlight interacts with a planetary surface, it can be reflected, absorbed, refracted, or diffracted, producing a return beam that is partially polarized. An important observable is the degree of polarization, which offers a diagnostic on the surface and sub-surface properties. The degree and orientation of this polarization depend on the physical characteristics of the surface, including grain size, porosity, and roughness, which together determine the balance between single and multiple scattering events (\textit{e.g.}, \citealt{Dollfus1985}). Past observations of airless bodies have highlighted two important effects:
\begin{enumerate}
    \item An increase of brightness at low phase angles, also called the brightness opposition effect (BOE)
    \item A negative linear polarization at small phase angles, also called the negative polarization branch (NPB)
\end{enumerate}

The linear degree of polarization $P(\alpha)$ is defined as:
\begin{equation}
    P(\alpha) = \frac{Q}{I} = \frac{I_{\perp} - I_{\parallel}}{I_{\perp} + I_{\parallel}},
\end{equation}
where $I_{\perp}$ and $I_{\parallel}$ are the light intensities measured perpendicular and parallel to the scattering plane (i.e., the Sun–object–observer plane), respectively. $\alpha$ is the phase angle, and $Q$ and $I$ are the second Stokes parameter and the total intensity, respectively. At low phase angles, the linear degree of polarization is negative (i.e., parallel to the scattering plane), while it becomes positive at larger phase angles (perpendicular to the scattering plane).

\begin{figure*}[h!]
    \centering
    \includegraphics[width=0.6\textwidth]{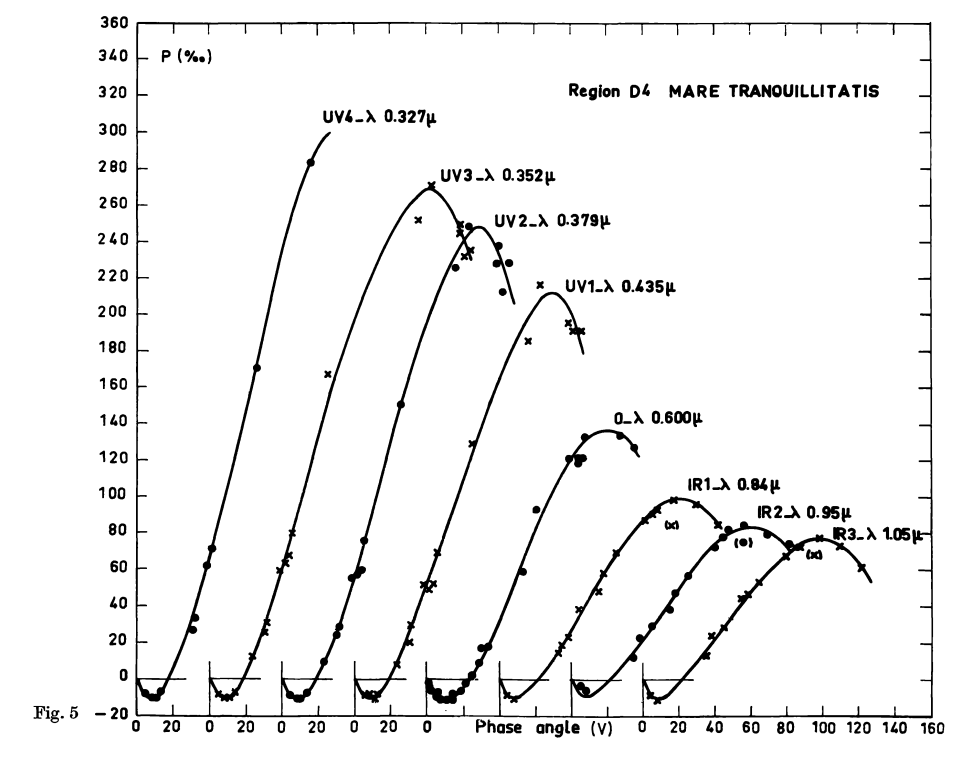}
    \caption{Polarization curves measured at eight wavelengths (from 327\,nm to 1.05\,$\micron$m) for \textit{Mare Tranquillitatis} on the lunar surface. From \citet{Dollfus1971}.}
    \label{fig:Dollfus1971}
\end{figure*}

While our interest lies in better understanding the surface properties of ocean worlds, it is worth briefly recalling the list of past polarimetric measurements performed over the lunar surface, and discussing to which extent they will be applicable to ocean worlds. An example of the degree of linear polarization measurements at different wavelengths on the lunar surface is shown on Fig. \ref{fig:Dollfus1971}, taken from \citet{Dollfus1971}.

At low phase angles, airless objects show a sharp increase in reflectance (BOE), and several optical mechanisms have been proposed to explain that phenomenon. Currently, a combination of (i) coherent backscattering, where backscattered light paths constructively interfere in the backward direction, (ii) mutual shadowing, and (iii) single-particle scattering explain the opposition phenomena (e.g., \citealt{Rosenbush2011}).

The second phenomenon (NPB) relates to the fact that $P(\alpha)$ becomes negative at phase angles lower than 20$^{\circ}$, reaching a minimum generally around 5–10$^{\circ}$, before crossing to positive polarization at the inversion angle (around $\sim$15–25$^{\circ}$). The negative polarization branch width and depth depend on surface structure. Fine, porous grains produce deeper minima, while compact or coarse surfaces yield weaker branches. A narrow negative polarization spike very close to $\alpha = 0^\circ$, seen in high-albedo objects, is attributed to coherent backscattering effects. Together, the BOE and NPB provide constraints on surface composition, roughness, and grain morphology \citep{Rosenbush2011}.

In the ultraviolet (UV), the interaction between light and matter is dominated by surface absorption and scattering within small grains. Polarization generally tends to increase toward shorter wavelengths (see, e.g., Fig. \ref{fig:Dollfus1971}), where absorption is stronger and multiple scattering less frequent. Since scattering events tend to randomize the polarization, which decreases with increasing wavelength \citep{Wolff1975}.

In the visible range (400–800\,nm), observations of the Moon and icy satellites show that polarization reaches a negative minimum at small phase angles, followed by an inversion (around 15–20$^{\circ}$) and a gradual increase to positive values at larger phase angles (see Fig.~\ref{fig:Dollfus1971}). The amplitude and wavelength dependence of the polarization curve depend strongly on albedo and particle size. 

In the near-infrared (NIR), absorption bands of H$_2$O ice, CO$_2$, salts and other volatiles strongly affect the reflectance spectrum in specific regions \citep{Trumbo2023, Villanueva2023, Cartwright2025_PSJ_JWST}. For instance, recent JWST observations on Europa revealed that CO$_2$ is concentrated within specific regions (chaos terrains), consistent with an endogenous carbon source, i.e., from the subsurface or deeper, rather than from an exogenic source. These studies also showed that exposed crystalline H$_2$O ice is similarly localized in these regions, pointing to active surface alteration due to diurnal subsolar heating. Modification of the structural properties of ocean world surfaces (e.g., crystalline vs amorphous ice, surface refreezing of impurities) is expected to show distinct polarimetric signatures that Pollux would measure. Laboratory measurements and remote-sensing observations show that polarization generally anti-correlates with albedo: bright, high single-scattering albedo surfaces polarize light only weakly, whereas darker, absorbing regoliths polarize more strongly (e.g., \citealt{Jeong2015}). 

\begin{figure*}[h!]
    \centering
    \includegraphics[width=0.7\textwidth]{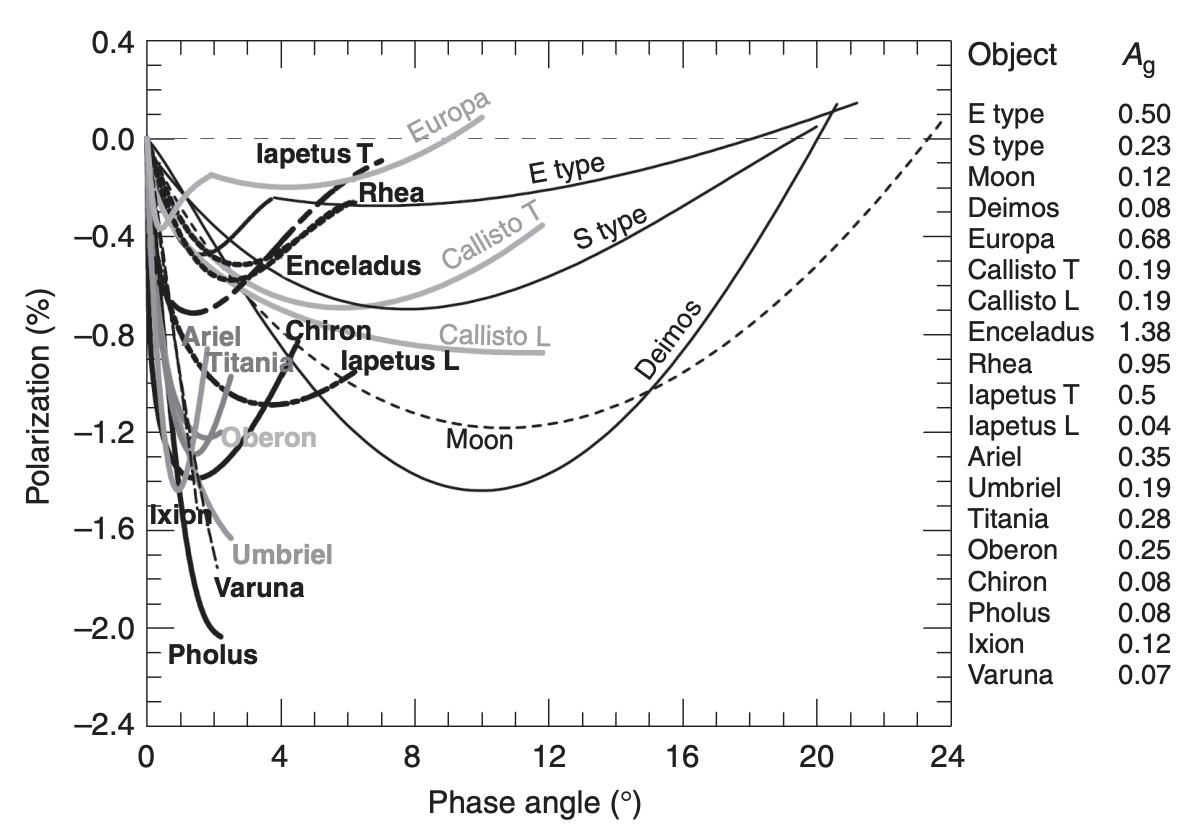}
    \caption{Comparison of polarization–phase curves for a range of Solar System bodies, including the Moon, Deimos, Europa, Callisto, Enceladus, Rhea, Iapetus, and the Uranian moons Ariel, Titania, Oberon, and Umbriel. For reference, E- and S-type asteroids, Centaurs (Chiron, Pholus), and trans-Neptunian objects (Ixion, Varuna) are also shown. The corresponding geometric albedos (A$_g$) are listed. The letters \textit{L} and \textit{T} corresponds to the leading and trailing hemisphere of given moons. From \citet{Rosenbush2015}, and references therein.}
    \label{fig:Rosenbush2015}
\end{figure*}

Within the jovian system, \citet{Dollfus1975} reported that Io and Europa, both characterized by high albedos compared to other small bodies or moons, display very low polarization, while darker Callisto is "polarimetrically quite different", exhibiting a well-defined negative polarization branch (see Fig.~\ref{fig:Rosenbush2015}). These studies also noted the dichotomy in the NPB between Callisto's trailing versus leading hemispheres, possibly linked with a dichotomy in its surface structural properties.

Fig.~\ref{fig:Rosenbush2015} presents the variety of the NPB measured on solar system bodies, observed using the 2.6-m and 1.25-m telescopes of the Crimean Astrophysical Observatory, and the 0.7-m telescope of the Chuguev Observation Station of the Institute of Astronomy of Kharkiv National University (see e.g., \citealt{Rosenbush2015} and references therein). In the case of high-albedo objects such as the Galilean moons, Saturn's rings and E-type asteroids, two minima have been reported within the NPB, with the more pronounced one located at phase angles lower than 2$^{\circ}$ \citep[see, e.g.,][and references therein]{Rosenbush2015}, and called the Polarization Opposition Effect (POE). More recent observations of the Io, Europa and Ganymede surfaces instead suggest that only the POE minimun exists within the NPB, and is attributed to coherent back-scattering of sunlight on microscopic surface icy grains \citep{Kiselev2022, Kiselev2024}.

Laboratory polarimetric measurements of various types of ice have been performed over the past few years by \citet{Poch2018}, providing a critical framework for interpreting polarimetric observations of ocean worlds. In their study, the polarimetric properties of water ice were investigated under controlled conditions by varying particle size, shape, and degree of thermal metamorphism. Their results show that the minimum of the NPB is primarily sensitive to the average particle size: as particle size decreases, the absolute value of the NPB minimum increases. They also found that freshly-condensed water vapor on cold surfaces produces oscillations in the degree of polarization, as part of the NPB. In addition, \citet{Poch2018} found that the polarization curves are extremely sensitive to the degree of thermal metamorphism experienced by the ice.

More generally, the wavelength dependence of both reflectance and polarization, spanning the ultraviolet to the infrared, provides a robust diagnostic of surface texture, composition, and internal structure. These signatures can be exploited by Pollux to characterize the structure and purity of water-ice surfaces, as well as to assess surface properties altered by contaminants, radiation processing, or mechanical compaction on ocean worlds throughout the Solar System. In particular, regions undergoing active surface processes, such as cryovolcanism, or sites associated with exchange between the surface and the subsurface ocean are expected to exhibit polarimetric properties that differ markedly from those of regions dominated by long-term, space-weathered ice.

\begin{figure*}[h!]
    \centering
    \includegraphics[width=0.6\textwidth]{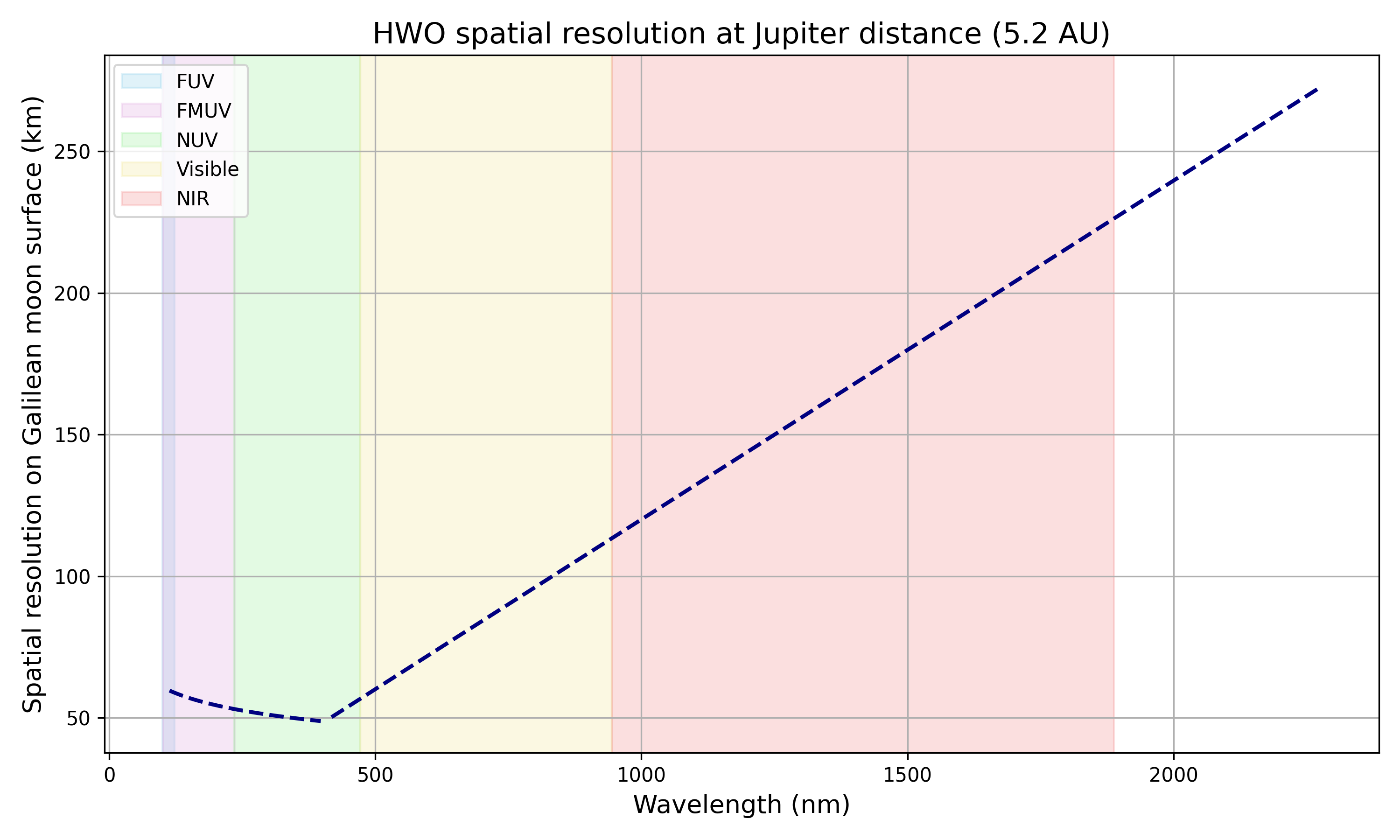}
    \caption{Spatial resolution of an 8-m diameter HWO telescope at the distance of Jupiter (5.2~AU) as a function of wavelength (Personal communication from K. France). The solid black curve shows the linear footprint on the surface. Colored shaded regions indicate the wavelength bands relevant for ultraviolet, visible, and near-infrared observations: FUV (100–123~nm), FMUV (101–236~nm), NUV (234–472~nm), visible (472–944~nm), and NIR (944–1888~nm). }
    \label{fig:HWO_res}
\end{figure*}

Figure~\ref{fig:HWO_res} illustrates the typical surface spatial resolution, expressed in kilometers, achievable on Europa with an 8\,m HWO-class telescope, over which these diagnostic quantities may be retrieved.

In summary, polarimetric observations from Pollux may provide information at sites of exchange with the interior ocean, where ocean-derived material may freeze differently or have composition-driven polarimetric differences. This would bring important information to better understand the composition and physical state of surface deposits, as well as their connection to material originating from the potential subsurface ocean.

\subsection{Ocean worlds electromagnetic environments and atmospheres}

Auroral emissions on the Galilean moons have been observed for several decades and reflect the interaction between the corotating jovian plasma and the moons’ local environments, modulated by their internal conductive layers (i.e., saline oceans), and Ganymede's intrinsic magnetic field \citep{Feldman2000, McGrath2013, Greathouse2022}. These moon auroral morphologies reflect the complex coupling between intrinsic or induced magnetic fields, tenuous atmospheres, and the jovian magnetospheric environment.

More specifically, HST revealed that Ganymede's aurorae are organized into two asymmetric oval structures tracing the open–closed field line boundary (OCFB) between Jupiter's and Ganymede's magnetic field lines \citep{Eviatar2001, Saur2015}. The auroral curtains are located at high latitudes on the trailing hemisphere (i.e., facing the flow of incoming jovian plasma) and lower latitudes on the leading side (i.e., in the wake of the plasma flow). Auroral structures were observed at the O\,\textsc{i} 130.4\,nm and 135.6\,nm lines \citep{Hall1998, Molyneux2018} driven by electron impact excitation of a tenuous, O$_2$-dominated atmosphere. 

The Juno mission, orbiting Jupiter since July 2016 \citep{Bolton2017}, revealed new details on Ganymede's aurora with Juno's UV spectrograph (UVS), which recorded its auroral curtains at spatial resolutions down to 4\,km, nearly 100 times finer than HST imagery, during perijove 34 and 35 \citep{Greathouse2022}. Juno-UVS revealed narrow, 25–50\,km-high curtains with sharp poleward boundaries. The morphology of the auroral ovals matched magnetic field models derived from Galileo and Juno magnetometer data, which confirms that their position coincides with Ganymede’s OCFB. Recent modeling of the Juno-UVS data measured precipitating electron fluxes of 0.1–2\,mW\,m$^{-2}$ with energies of 20–300\,eV \citep{Benmahi2025}.

The unprecedented spatial resolution of Juno-UVS over these emissions also revealed the presence of small-scale auroral sub-structures in the form of localized patches, primarily emitting in the O I 130.4 and 135.6 nm lines produced by electron precipitation. These auroral patches, with typical sizes of $\sim$50 km and brightnesses reaching $\sim$200 Rayleigh, are observed on the leading, downstream side of Ganymede and map approximately to the downstream reconnection region. Their morphology and location closely resemble terrestrial auroral beads associated with substorms and jovian mesoscale aurora observed prior to dawn storms, suggesting that similar magnetospheric processes may operate across planets and moons despite their differing plasma environments and dynamical regimes \citep{Moirano2026}.

The airglow emissions at Europa and Callisto contrast with those observed at Ganymede. The electrons responsible for exciting these emissions are thought to differ in nature, in that they may not be specifically accelerated. For Europa, it has been shown that the thermal electron population from the plasma torus may be sufficient to excite the emissions, producing brightnesses on the order of those observed at Ganymede, i.e., in the 50–100\,R range for the combined O I 130.4 and 135.6 nm lines. The polarimetric characteristics of the auroral signal are therefore expected to differ from Ganymede's, although the extent of this difference remains unclear. Thermal electrons are expected to have preferred directions, primarily along the magnetic field $B$, and may thus result in polarized emission. However, uncertainties in both atmospheric and electron properties prevent definitive conclusions regarding the generation of the aurora, consequently, measurements of the polarization state would provide crucial additional constraints \citep{Hall1995, Roth2016, Roth2021}. Similarly, for Callisto, the excitation mechanisms responsible for the emissions remain poorly constrained, with photoelectrons or magnetospheric electrons suggested as possible sources \citep{Cunningham2015, DeKleer2023}.

At Earth, the polarization of auroral emissions provides valuable insight into the anisotropy of precipitating electron populations. The degree of linear polarization for both the red \citep{lilensten2008} and green \citep{Bosse2020, Bosse2022} lines were observed. These polarimetric measurements allow for further derivation of information on ionospheric electrodynamics, such as the characteristics of electron precipitation and the morphology and dynamics of E-region currents. \citet{Bommier2011} developed the first theoretical framework describing impact polarization of the 630\,nm red auroral line, which was later applied by \citet{Lilensten2015} using a self-consistent electron transport model. Their calculations predict degrees of linear polarization ranging from 0.6 to 1.8\%, depending on altitude and line-of-sight integration, in good agreement with ground-based observations from Svalbard ($1.9\%\pm0.1\%$). 

This polarization arises from anisotropic excitation by magnetospheric electrons, with the polarization orientation depending on the characteristic electron energies. The relatively weak polarization levels result from partial depolarization due to collisions and the contribution of competing isotropic excitation sources. \citet{Lilensten2015} further showed that the polarization signal is highly sensitive to several parameters, including the electron scattering function and the magnetic field topology.

Despite multiple observational campaigns at Earth, significant uncertainty remains regarding the polarization of auroral lines. More specifically, no polarization of the terrestrial auroral green line is expected \citep{Bosse2022}, based on the quantum mechanical calculations of \cite{Bommier2011}. Space-based observations of auroral emission polarization would prevent any possible contamination from polarization due to scattered light. Spectropolarimetric measurements of auroral emissions at ocean worlds have not been previously explored, and could provide a new and indirect diagnostic of the properties and dynamics of precipitating magnetospheric electron populations.

Unlike all the other outer planet moons, Titan possesses a dense atmosphere with a persistent photochemical haze, making it an ideal target for polarimetric measurements to constrain aerosol microphysics. Spacecraft polarimetric observations recorded by Pioneer 11 and Voyager showed high degrees of linear polarization in the blue and lower values at red wavelengths. These measurements are consistent with a bimodal haze particle distribution that includes non-spherical particles, to reproduce both the strong forward scattering and the high degree of single-scattering polarization \citep[see, e.g.,][and references therein]{Bagnulo2024}.

More recently, HST imaging polarimetry from the UV to the NIR revealed negligible polarization at Titan's disc center and a strong increase toward the limb, a signature likely due to Rayleigh-like scattering processes \citep{Lorenz2004,Lorenz2006}. Pollux spectropolarimetry would extend these constraints into the UV with high spectral resolution, enabling sensitive tests of haze particle size, morphology, and distribution in atmospheres and exospheres in the outer Solar System.

\section{Summary and Conclusion}

Additional polarimetric measurements of ocean world surfaces remain rather limited. Ground-based high-contrast imaging polarimeters such as ZIMPOL (Zurich Imaging Polarimeter), part of the SPHERE instrument, operate in the 500–900\,nm spectral range and provide high spatial resolution and high-contrast imaging polarimetry in the immediate surroundings of bright stars \citep{Schmid2018}. Resolved spectro-polarimetric datasets of the ocean worlds surface are currently sparse, though the adaptive optics of SPHERE/ZIMPOL would allow deriving details on the order of 100\,km at the surface of Europa \citep{Patty2023Polarimetry}.

Pollux, with its UV–NIR spectropolarimetric capabilities, would provide a unique opportunity to advance the study of ocean worlds in the Solar System. By simultaneously measuring both spectral and polarization signatures, Pollux will enable precise characterization of the scattering properties of icy moon surfaces and atmospheres, as well as their airglow emissions. Polarimetric data will allow the derivation of key surface parameters such as grain size, porosity, composition, as well as atmospheric haze particle size, morphology, and distribution. Additionally, the detection of auroral polarization would open new diagnostics for magnetospheric particle interactions with moon exospheres.

Pollux will provide unprecedented capabilities beyond what HST and JWST currently offer in space-based UV spectro-polarimetric observations of icy moons, delivering important insights into their surface-interior interactions within the Habitable Worlds Observatory framework.

{\bf Acknowledgements.} V. Hue and B. Benmahi acknowledge support from the French government under the France 2030 investment plan, as part of the Initiative d’Excellence d’Aix-Marseille Université – A*MIDEX AMX-22-CPJ-04, as well as support of CNES to the Juno and JUICE missions.

\bibliography{author.bib}

\end{document}